# Ω FROM VELOCITIES IN VOIDS

**Avishai Dekel**[1] and **Martin J. Rees**[2]

[1] Racah Institute of Physics, The Hebrew University, Jerusalem 91904, Israel
dekel@vms.huji.ac.il
[2] Institute of Astronomy, Madingley Road, Cambridge CB3 0HA, UK
mjr@mail.ast.cam.ac.uk

## ABSTRACT

We propose a method for deriving a dynamical lower bound on Ω from diverging flows in low-density regions, based on the fact that large outflows are not expected in a low-Ω universe. The velocities are assumed to be induced by gravity from small initial fluctuations, but no assumptions are made regarding their exact Gaussian nature, galaxy biasing, or Λ. The derivatives of a diverging velocity field infer a nonlinear approximation to the mass density, which is an overestimate when the true value of Ω is assumed. This inferred density can become ridiculously negative when the assumed Ω is too low, thus bounding Ω. Observed radial peculiar velocities of galaxies allow the POTENT procedure to recover the required velocity field, Gaussian smoothed at 1200 km s$^{-1}$. The density and the associated errors are then inferred for different values of Ω, searching for a void which shows negative inferred density with high confidence. A preliminary application to data in a southern void indicates that $\Omega \leq 0.3$ can be ruled out at the 2.4-sigma level. A detailed study of possible systematic errors is under way.

*Subject headings:* cosmology — dark matter — galaxies: clustering — galaxies: formation — gravity — large scale structure of universe

# 1. INTRODUCTION

The observed large-scale peculiar velocities can place dynamical constraints on the cosmological density parameter, $\Omega$, in several different ways. First, if one is willing to assume an explicit *"biasing"* relation between galaxy and mass density (see Dekel & Rees 1987 for a review), one can estimate $\Omega$ by comparing the peculiar velocity data to a uniform galaxy redshift survey. Using linear gravitational instability theory, and assuming a linear biasing relation with a factor $b$ between the smoothed density fluctuation fields of galaxies and mass, one can measure the ratio $\Omega^{0.6}/b$ (*e.g.* Dekel *et al.* 1993; Hudson *et al.* 1994). The degeneracy between $\Omega$ and $b$ could, in principle, be removed by nonlinear analysis, but the interpretation is always confused by possible nonlinearities in the biasing relation, leaving the $\Omega$ determination strongly dependent on the galaxy biasing scheme assumed.

Alternatively, one can determine $\Omega$ from the velocity field alone, if one is willing to assume a specific probability distribution function (*PDF*) for the initial density fluctuations, *e.g.* Gaussian. This can be done by tracing the observed velocity field back in time, and comparing the recovered initial *PDF*, which depends on $\Omega$, with the assumed initial *PDF* (Nusser & Dekel 1993). It can also be done by comparing the measured *PDF* of the velocity-divergence with the predictions for a system which evolved from Gaussian initial conditions under gravity with an assumed $\Omega$ (Bernardeau *et al.* 1994).

Here, we propose a method that relies *only* on the assumptions that the large-scale structure evolved under *gravity* from small-amplitude initial fluctuations, and that the galaxies are honest tracers of the underlying velocity field. The unique feature of this method is that it requires no assumptions about how the galaxies are distributed relative to the mass, or concerning the specific *PDF* of the initial fluctuations.

The intuitive idea underlying our method is that streaming motions arise from the "pull" of overdense regions and the "push" of underdense regions. In a low-$\Omega$ universe one would not expect substantial relative deviations from the Hubble flow except for converging flows around very overdense regions, which only fill a small fraction of the volume. In particular, one can easily see that there is an $\Omega$-dependent limit on the magnitude of the diverging flow around a void: even a completely empty void cannot induce substantial peculiar motions compared to the Hubble velocity if the mean density around it is well below the critical density.

For a given diverging velocity field, and a given value of $\Omega$, one can infer from the partial velocity derivatives, $\partial \boldsymbol{v}/\partial \boldsymbol{x}$, nonlinear approximations with small variance to the mass-density fluctuation field, $\delta \equiv \bar{\rho}/\rho - 1$ (Nusser *et al.* 1991, hereafter NDBB). One such approximation, $\delta_c$, has been found to overestimate the true density in density wells ($\delta_c > \delta$), under very general conditions, as long as the assumed value of $\Omega$ is the true value. We use this approximation below.

Analogously to the behavior of the linear approximation, $\delta_0 = -\Omega^{-0.6} \boldsymbol{\nabla} \cdot \boldsymbol{v}$, the $\delta_c$ inferred from a given velocity field is a monotonically increasing function of $\Omega$, and it may become smaller than $-1$ for $\Omega$ values that are too small below the true value. Then, since $\delta \geq -1$ because mass is never negative, the allowed values for $\Omega$ are bounded from below.

Given the observed radial peculiar velocities of galaxies, one can use the POTENT procedure to recover the three-dimensional velocity field, Gaussian smoothed at $\sim$ 1000 km s$^{-1}$ (Dekel *et al.* 1990, DBF; Bertschinger *et al.* 1990, BDFDB; Dekel *et al.* 1994). One can then derive the inferred mass-density field and the associated *error* field for different values of $\Omega$. Focusing on the deepest density wells, the assumed $\Omega$ should be lowered until $\delta_c$ becomes significantly smaller than $-1$, where the significance is estimated based on the estimated errors. Such low values of $\Omega$ could be ruled out.

When trying to apply this test to the available data, severe difficulties are posed by the large random and systematic uncertainties due to distance errors and nonuniform, sparse sampling. In our effort to assign statistical significance to the lower bound on $\Omega$, we can rely on the random error estimates of POTENT based on Monte Carlo noise simulations. The systematic errors have been partially corrected for in the POTENT procedure, but a more specific investigation of the biases affecting the smoothed velocity field in the deepest density wells is required for more reliable results.

The method is outlined in §2. The preliminary results from current data are briefly described in §3. The analysis is discussed in §4.

## 2. METHOD

The raw data consist of a large set of galaxies with measured redshifts $z_i$ and redshift-independent estimated distances $r_i$ of random Gaussian errors $\sigma_i \simeq (0.15-0.21)r_i$. The peculiar velocities are $v_{ri} = z_i - r_i$. The distances are corrected for inhomogeneous Malmquist bias by heavy grouping (Willick *et al.* 1994) and by adopting a correction procedure which assumes that IRAS galaxies (Fisher 1992) trace the underlying galaxy distribution (Dekel *et al.* 1994). In POTENT, the radial peculiar velocities are first smoothed with a Gaussian window of radius 1200 km s$^{-1}$ onto a spherical grid, yielding a radial field $v_r(\boldsymbol{x})$. The smoothing suppresses the random uncertainties due to distance errors and sparse, nonuniform sampling. The adopted weighting scheme attempts to minimize the associated systematic biases. POTENT then recovers the transverse components of the velocity field by requiring that it is derived from a scalar potential, $\boldsymbol{v} = -\boldsymbol{\nabla}\phi$, as predicted by linear gravitational instability and maintained later via Kelvin's circulation theorem under appropriate smoothing (Bertschinger & Dekel 1989). The potential at each point $\boldsymbol{x}$ is computed by integrating $\phi(\boldsymbol{x}) = -\int_0^r v_r(r', \vartheta, \varphi) dr'$, and the transverse velocity components are derived by differentiating this potential.

Given the three-dimensional velocity field and $\Omega$, one can approximate the corresponding mass-density fluctuation field. In the linear approximation, the density is $\delta_0 = f(\Omega)^{-1}\boldsymbol{\nabla}\cdot\boldsymbol{v}$, where $f(\Omega) = \dot{D}/D \approx \Omega^{0.6}$, and $D(t)$ is the fluctuation linear growth factor (Peebles 1980). The velocities are measured in km s$^{-1}$, which eliminates any dependence on the Hubble constant. Although $\delta_0$ is also an exact solution to the nonlinear Euler and Poisson equations for cosmological dust under the quasilinear approximation of a universal time dependence for the displacements (Zel'dovich 1970), it always underestimates $\delta$ in the nonlinear regime (NDBB). Therefore, the linear approximation *cannot* be used to obtain the desired lower limit on $\Omega$.

A better quasilinear approximation is the solution to the continuity equation assuming Zel'dovich displacements,

$$\delta_c(\bm{x}) = \left\| \bm{I} - \frac{1}{f(\Omega)} \frac{\partial \bm{v}}{\partial \bm{x}} \right\| - 1 , \tag{1}$$

where $\bm{I}$ is the unit matrix, the derivatives are taken with respect to the Eulerian comoving position, and the double vertical bars denote the Jacobian determinant (NDBB). (This is the standard approximation currently used in the POTENT procedure.) The desired inequality, $\delta_c \geq \delta$ at the negative tail, can be shown analytically to hold in a spherical negative top-hat model (NDBB, Fig. 1a). This inequality is *generic*, for any value of $\Omega$ in the range of interest, as long as the structure has grown according to the conventional wisdom. This is illustrated in Figure 1, which shows the mean and standard deviation of $\delta_c - \delta$ at a given true $\delta$ in N-body simulations of $\Omega = 1$ and $0.2$. These simulations started off as random Gaussian realizations of the CDM power spectrum in a box of $25{,}600$ km s$^{-1}$, with $128^3$ grid points and particles. They were evolved using a PM code until $b = 1$. The final density and velocity fields were smoothed with a Gaussian window of radius $500$ km s$^{-1}$. The moments of $\delta_c - \delta$ are computed in bins of width $0.1$ in $\delta$. The figure shows that at the negative tail, $\delta_c$ is many standard deviations larger than $\delta$ (see Mancinelli et al. 1994 for more details). This behavior is not specific to the assumed fluctuation power-spectrum; a similar behavior is exhibited, for example, by HDM simulations (NDBB). The behavior of the mean is also similar when a larger smoothing is applied; for $1200$ km s$^{-1}$ smoothing the standard deviations get smaller by about a factor of three, and the deepest void is of $\delta \approx -0.7$ only.

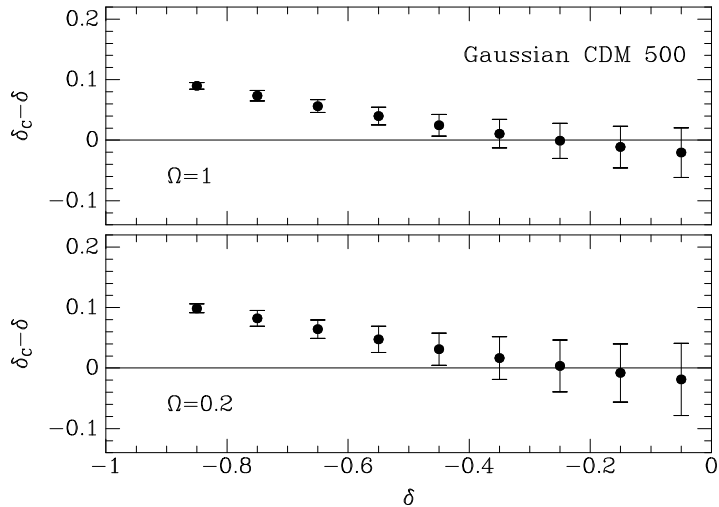

**Figure 1:** The approximation inferred from the velocity derivatives, $\delta_c$, compared with the true density fluctuation, $\delta$, in a Gaussian CDM N-body simulation. The fields are smoothed at $500$ km s$^{-1}$. Only negative fluctuations are shown, focusing on the upward deviation of the mean at the negative tail, and the small standard deviation. This behavior is robust: it is valid for a wide range of $\Omega$ values and under general initial conditions.

While the inequality $\delta_c \geq \delta$ is necessary for the method to work, the discriminatory power of the test could be improved by using an approximation which overestimates $\delta$ at the negative tail in the minimal possible way.

The main source of uncertainty in the $\delta_c$ field such derived by POTENT is the random distance errors. To estimate this uncertainty we construct Monte-Carlo samples in which the peculiar velocity of each galaxy is perturbed by a Gaussian random variable of dispersion $\sigma_i$. For each perturbed sample, we compute $\delta_c$ at each grid point. It's standard deviation over the Monte Carlo simulations, $\sigma(\boldsymbol{x})$, serves as our error estimate for the purpose of evaluating the significance of the limit on $\Omega$.

The bias arising in the smoothing procedure due to sampling gradients remains severe in very poorly-sampled regions despite our efforts to minimize it by appropriate weighting. We therefore exclude regions where the sampling is unacceptably poor, by considering only grid points where the distance to the 4th nearest neighboring object in the data set, $R_4$, is smaller than a critical value, chosen here to be 1500 $\mathrm{km\,s^{-1}}$. Such regions typically occur in the current data at large radii beyond 6000 $\mathrm{km\,s^{-1}}$, and in the galactic zone of avoidance.

## 3. PRELIMINARY RESULTS

As an illustration for the use of the proposed test, we apply it to the tentative data recently compiled by Willick *et al.* (1994), and POTENT analyzed by Dekel *et al.* (1994). It is based on $\sim 3000$ galaxies, mostly spread inside a sphere of radius 6000 $\mathrm{km\,s^{-1}}$ about the Local Group, and extending beyond 8000 $\mathrm{km\,s^{-1}}$ in selected regions.

The most promising "test case" provided by this data set seems to be a broad diverging region centered near the supergalactic plane at the vicinity of $(X, Y) = (-2500, -4000)$ in $\mathrm{km\,s^{-1}}$ supergalactic coordinates. This region can be roughly identified with the "Sculpter void" of galaxies (Kauffman *et al.* 1991), clearly seen in the SSRS redshift catalog next to the "Southern Wall" (daCosta *et al.* 1988, middle of figure 7b).

Figure 2 shows maps of $\delta_c$ in the relevant part of the supergalactic plane, for several different values of $\Omega$. The void is confined by the Pavo part of the Great Attractor on the left, the Aquarius Wall extension of the Perseus-Pisces supercluster on the right, and the Cetus Wall from the bottom at a larger distance (not seen in the plot). Values of $\Omega \approx 1$ are perfectly consistent with the data, but $\delta_c$ becomes smaller than $-1$ in this void already for $\Omega = 0.6$.

The confidence by which $\delta_c < -1$ at a given point is expressed in terms of the random error $\sigma$ there. The values $\Omega = 0.4, 0.3$, and $0.2$ are ruled out at the 1.6-, 2.4-, and 2.9-sigma levels respectively.

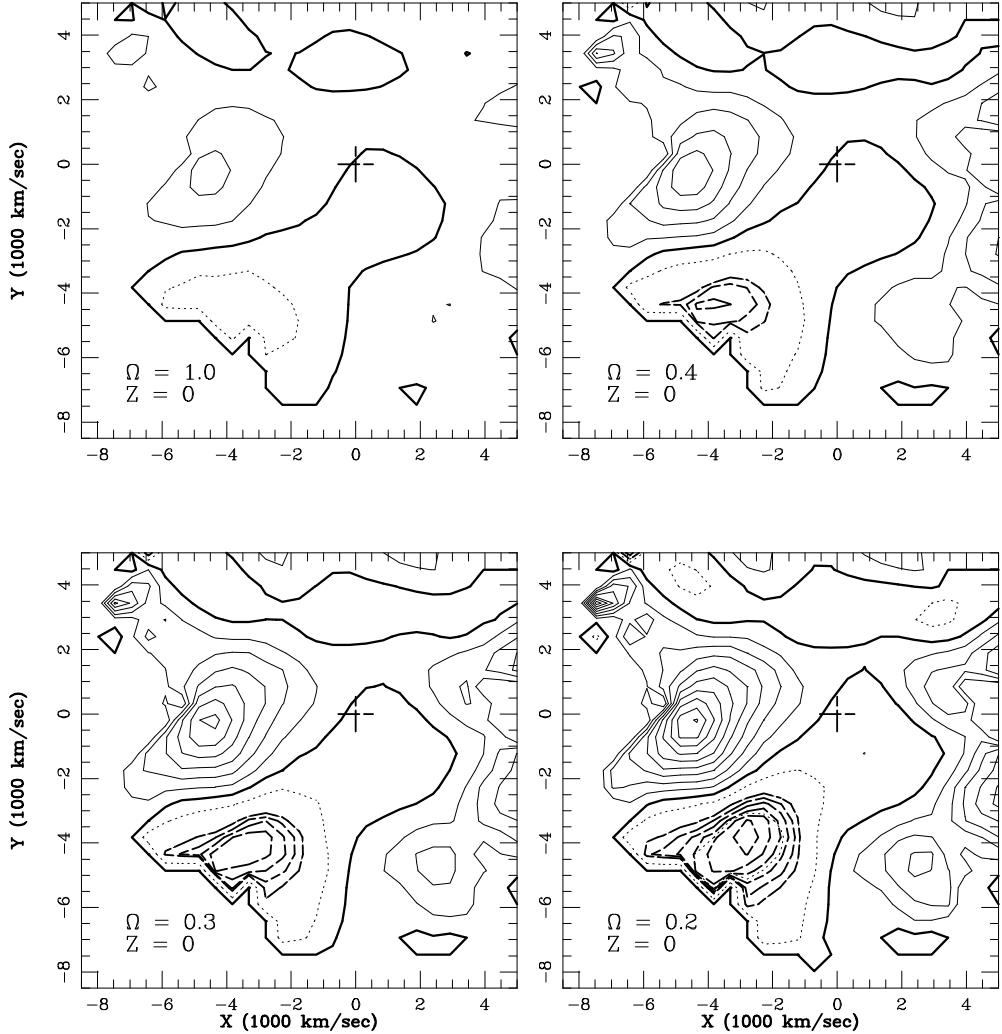

**Figure 2:** Maps of the density-fluctuation field inferred from the observed velocity derivatives, $\delta_c$, in part of the supergalactic plane, for several different values assumed for $\Omega$. The void of interest is confined by the Pavo part of the Great Attractor on the left and the Aquarius extension of the Perseus-Pisces supercluster on the right. The Local Group is marked by a '+'. Contour spacing is 0.5, with the mean, $\delta_c = 0$, marked by a heavy line, $\delta_c > 0$ solid, and $\delta_c < 0$ dotted. The heavy-dashed contours mark the downward deviation of $\delta_c$ from $-1$ in units of the standard deviation $\sigma$, starting from zero (which coincides with $\delta_c = -1$), and decreasing with spacing $-0.5\sigma$. The value $\Omega = 0.3$ is ruled out at the 2.4-sigma level.

## 4. DISCUSSION

We are proposing a dynamical method for determining a robust lower bound on $\Omega$ from observed diverging peculiar velocities. Our purpose in the present note is to present the basic method and to report the encouraging preliminary result of its application to data, in order to motivate further observations and trigger further tests and improvements.

Our preliminary result indicates that the method has the potential of ruling out low values of $\Omega$. Based on the available data and the current bias-correction procedure of POTENT, the most pronounced diverging flow in our cosmological neighborhood indicates that $\Omega = 0.3$ or smaller can be ruled out at the $> 2.4$-sigma level. Results of higher confidence should await a more careful study of the possible systematic errors.

Taking the current result at face value, it is consistent with other recent dynamical lower bounds to $\Omega$ from observations on large scales. By comparing the density fields of IRAS galaxies and POTENT mass, Dekel et al. (1993) find $f(\Omega)/b_{iras} = 1.3^{+0.8}_{-0.6}$. A similar comparison between POTENT mass and optical galaxies yields $f(\Omega)/b_{optical} \approx 0.65 \pm 0.2$ (Hudson et al. 1994). By demanding Gaussian initial PDF (e.g. based on Nusser, Dekel & Yahil 1994), Nusser & Dekel (1993) find that $\Omega \leq 0.3$ can be rejected at the 6-sigma level. A similarly high value for $\Omega$ is indicated by the skewness of $\nabla \cdot \boldsymbol{v}$, under the same assumption of Gaussian initial PDF (Bernardeau et al. 1994).

Having mentioned the preliminary results, the following discussion focuses on certain features of the *method*.

The bound refers to the mass-density parameter $\Omega$ and is insensitive to the value of the cosmological constant $\Lambda$; it is actually a bound on Peebles' $f(\Omega)$ which is hardly affected by $\Lambda$.

The method is *independent* of the biasing relation between galaxy and mass density fields; galaxies serve only as tracers of the gravitationally-induced velocity field. It therefore bypasses the complexities and uncertainties of how galaxies form and how their properties may depend on their environment.

The main leverage which enables us to constrain $\Omega$, either by the PDF methods or by the void method of the present letter, is the extreme sensitivity of the recovered density in low-density regions to the assumed $\Omega$ when it is small.

The method does not require any special topology for the isodensity contours associated with the diverging flow; it does not have to be an isolated spherical void. The POTENT procedure, which translates the data into a full velocity field, allows the computation of the key quantity $\delta_c(\partial \boldsymbol{v}/\partial \boldsymbol{x}, \Omega)$ wherever there are enough tracers.

The relation between the velocities and $\Omega$ assumes that the mass inhomogeneities evolved via gravitational instability from small initial fluctuations. The necessary properties of potential flow and the inequality $\delta_c > \delta$ have been tested to be valid for the smoothed fields in simulated gravitating systems with *Gaussian* initial fluctuations and "standard" power spectra (BD; NDBB). However, these properties are found to be valid under more general initial conditions; e.g. in explosion scenarios where the initial conditions are neither Gaussian nor of gravitational origin (Babul et al. 1994). We also know that these properties are valid in a spherical uniform void in which the velocity field is properly sampled interior to the caustic edge. Thus, the method does not rely explicitly on the specific statistical properties of the initial fluctuations. How peculiar and different from a random Gaussian field can the initial fluctuations be before the method possibly breaks down is yet to be investigated. One could probably invoke contrived, extremely non-Gaussian counter examples in a low-$\Omega$ universe, where the effective $\Omega$ in an extended diverging region is much larger (*e.g.* in an average-density region surrounded by several big clusters).

A disadvantage of the proposed method relative to the methods which use the whole *PDF* is that it makes use of only a limited portion of the available data — the bound is effectively based on the velocities near the deepest nearby void.

Two systematic effects deserve special attention. First, the sparse and nonuniform sampling. Although the method could be applied even if there were no correlation between the distribution of galaxies and the dark matter density, such correlation does exist, making it hard to trace the velocity field in the low-density regions. The smoothed velocity field at the center of a void could be biased by the interpolation from more populated regions surrounding it, introducing a "sampling-gradient bias" (SG, see DBF). The *POTENT* method tries to correct for this bias, and we exclude regions of extremely poor sampling where the correction fails, but even in the best possible void one may be left with certain SG bias which should be accounted for in detail.

The second effect, the inhomogeneous Malmquist bias (IM, see Willick 1994), is the most fearsome enemy of all the dynamical estimates of $\Omega$ mentioned above. The random distance errors, combined with galaxy-density gradients along the line of sight, systematically enhance the inferred velocity gradients and the inferred density fluctuations. The IM bias is of particular relevance here in view of the possibly large log-density gradients near the void edges. This bias is always towards overestimating $\Omega$. In the current *POTENT* analysis we correct for IM bias using a procedure which has been tested to work properly in general, but there is still room for improvement, with particular attention to the IM bias in voids.

The two systematic uncertainties just discussed are being carefully studied using simulations, with the aim to refine and calibrate the method (Ganon and Dekel 1994). But while this theoretical effort is going on, it would be of great importance to pursue a parallel observational effort to improve the sampling of peculiar velocities in low-density regions in order to reduce the errors. The void tentatively addressed above is an obvious place to start such an exploration, but it is certainly not the only void of possible relevance. For the method to be effective we need to find a void that is (a) bigger than the correlation length for its vicinity to represent the universal $\Omega$, (b) deep enough for the lower bound to be strong, (c) nearby enough for the distance errors to be small, and (d) properly sampled to trace the velocity field in its vicinity.

## Acknowledgments

We thank G. Ganon and T. Kolatt for their contribution. We acknowledge AD's *POTENT* collaborators, D. Burstein, S. Courteau, A. Dressler, S.M. Faber, J. Willick, and A. Yahil, for the preliminary *POTENT* results used to test the feasibility of the method presented here. AD is grateful for the hospitality of the Institute of Astronomy at Cambridge where certain parts of this work were done. This research has been partially supported by the Israeli National Science Foundation, and by the US-Israel Binational Science Foundation.